# Privacy Impacts of Data Encryption on the Efficiency of Digital Forensics Technology

Adedayo M. Balogun

School of Computing and Mathematics,
University of Derby,
Derby, United Kingdom.

Shao Ying Zhu

School of Computing and Mathematics,
University of Derby,
Derby, United Kingdom.

*Abstract*—Owing to a number of reasons, the deployment of encryption solutions are beginning to be ubiquitous at both organizational and individual levels. The most emphasized reason is the necessity to ensure confidentiality of privileged information. Unfortunately, it is also popular as cyber-criminals' escape route from the grasp of digital forensic investigations. The direct encryption of data or indirect encryption of storage devices, more often than not, prevents access to such information contained therein. This consequently leaves the forensics investigation team, and subsequently the prosecution, little or no evidence to work with, in sixty percent of such cases. However, it is unthinkable to jeopardize the successes brought by encryption technology to information security, in favour of digital forensics technology. This paper examines what data encryption contributes to information security, and then highlights its contributions to digital forensics of disk drives. The paper also discusses the available ways and tools, in digital forensics, to get around the problems constituted by encryption. A particular attention is paid to the *Truecrypt* encryption solution to illustrate ideas being discussed. It then compares encryption's contributions in both realms, to justify the need for introduction of new technologies to forensically defeat data encryption as the only solution, whilst maintaining the privacy goal of users.

## I. INTRODUCTION

Data is becoming largely existent in today's world than they were anticipated some three decades ago [1]. Individuals are keeping lot more amount of information than organizations kept in the yesteryears. Significant amounts of such information are valued and consequently preferred to be known to them alone. Such valued information includes their financial details, medical records, locations, as well as professional and network information. Businesses and organizations possess larger amounts of information than individuals. A good amount of such information is critical to their sustained existence and growth. Their intellectual properties and trade secrets are kept away from potential exploits, thus, considered very private. Governments and agencies keep sensitive information that may affect the stability of their jurisdictions, politically or economically, if divulged.

The necessity to keep such information within the required confines describes a component purpose of Information Security, which involves the totality of activities to ensure the protection of information assets that use, store, or transmit information from risk through the application of policies, education, training, awareness, and technology [2]. Data security involves the consideration of potential confidentiality, integrity, and availability threats to data services, using functions such as identification, authentication, authorization and audit [3].

An important and popular methodology for enforcing information security is encryption, which is itself an element of cryptography. Cryptography provides a secret communication mechanism between two or more parties. Symmetric and Public Key Cryptography employ various algorithms to ensure the security of data items 'at rest', 'in use', and 'in motion'.

Data encryption may not be an explicit solution to information security problems, as organizations remain increasingly vulnerable to data breach incidents, but it is still the most efficient fix when deployed adequately [4]. This has led to the growing availability of full disk encryption tools. Disk manufacturers are embedding full encryption tools into their products, making encryption more available for use [5].

The study conducted by [4] showed the increased usage of full disk, virtual volume, native disk, and flash drive encryptions over two years. However, for reasons other than the cost of deployment and managing an encryption solution, some organizations have shunned or still undecided about adopting encryption solutions. They insisted that "availability is more important than confidentiality" [6]. The time the encryption and decryption processes take before data is made accessible to potential users may cause delay in organizations' operations, depending on how complex the base algorithm is. Such delays may escalate to a sort of denial-of-service situation, which may be adverse to organizations' businesses. On the electronic discovery front, unavailability problem prevents anticipated investigation of cyber-incidents [7]

The rest of this paper is organized thus: Section 2 presents the reason for this work. Section 3 highlights the contributions of data encryption to information security and digital forensics. Section 4 discusses the effects of data encryption on digital forensics processes, as well as the currently improvised digital forensics methods to defeat data encryption issues. In section 5, justifications for new technologies to help forensic investigation of encrypted data containers are discussed. Conclusions are given is Section 6.

School of Computing and Mathematics, University of Derby, United

## II. NECESSITY & SCOPE

Surveys revealed the continuously increasing adoption of cryptographic solutions by organizations for various data security platforms within the last five years [6]. The report of the surveys infers the anticipation of non-users to adopt partial or holistic cryptographic solutions in the nearest future. This suggests the impending domination by cryptographic procedures, to protect information in the computer world. There are ways for investigators to outmaneuver the use of cryptography as a provocation to digital forensics processes. These methods are either by legally obtaining appropriate 'search and seize' authorizations or tactically planning to catch the offender unawares and hence, access live – running and unencrypted – systems [5]. However, only a handful of encryption incidents encountered by investigators have been solved using those methods. The larger lot of about 60% often does not get prosecuted, not because they were missed, but because nothing could be done to access the potential evidence [8][9]. An instance is the case of Brazilian banker Daniel Dantas, whose strong truecrypt passphrase has foiled all attempts by Brazilian police and FBI to access his encrypted potential evidential hard drives [10][11].

The inconsistency of legal systems across boundaries does not make the process easier, as laws may or may not enjoin perpetrators to help the investigators access the encrypted medium [12]. This is evident in the Dantas' suspected money laundering case, where Brazil had no legislation to make him reveal his passphrase or encryption type, unlike the United Kingdom [11]. Therefore, researchers and developers need to be reminded of privacy-enforcement threats to forensic investigations, and pestered about the need for technologies to help deal with accessing encrypted storage devices.

## III. ENCRYPTION CONTRIBUTIONS

### A. Data Encryption for Information Security

In order to examine threats contributed by a technology, the solutions it offers should be considered too [13]. Encryption, as an element of cryptography, is a methodology for achieving information security, through secretive communications [14].

The United Kingdom's Data Protection Act 1998 most suitably describes the confidentiality element of information security. It seeks to ensure that the information held by organizations of their customers and employees are safeguarded from other uses than they were obtained [15]. This is meant to avert incidents such as identity crimes, and protect such potential victims from damages and embarrassment that unauthorized use of their data may cause [16]. The powers conferred on the Information Commissioner's Office (ICO) and the Financial Services Authority (FSA) to spot check and fine defaulting organizations, as well as the necessity for card-accepting organizations to comply with industry standards, like the Payment Card Industry Data Security Standard (PCI DSS), has led to the increasing adoption of encryption solutions [6].

There is also a huge necessity to ensure the confidentiality of data items, at rest, in use, or in motion [17]. Financial organizations, where transactions are regularly performed on data, have to ensure that such data are not subject to unauthorized access or modifications. The combination of the encryption and hash technologies to create digital signatures and certificates, which are used to ensure data confidentiality and integrity, is a laudable approach [18].

As far as information security is concerned, data encryption technology has been of invaluable success on the confidentiality and integrity fronts. Whereas on the availability front, it is known for delays on sparse occasions. Serious availability issues caused by the deployment of encryption solutions are not unheard of, although they are usually addressable by providers [19]. In an overall sense, it is hence, agreeable to regard data encryption as a massive solution for information security challenges.

### B. Data Encryption for Digital Forensics

During their hard disk sanitization study at Massachusetts Institute of Technology's Laboratory, [14] found out the ease with which data can be retrieved from disk drives. The ability to recover deleted data and locate hidden data was not a challenge, because there were forensic tools that require little or no specialized user training in existence. The success of those tools was attributed to the "widespread failure of the market to adopt encryption technology" [14]. Some eight years after his work at the MIT laboratory, [20] admitted that the current forensic tools are struggling to be useful to digital forensics investigators when certain data are concerned. He stressed the increasing occurrence of such data and identified format incompatibilities, encryption and lack of training as the reasons. [21] Highlighted data scalability and encryption as some of the unaddressed issues too.

Encryption of data on disk drives is implemented at the file system encryption and full disk encryption (bitstream) levels. At the file system encryption level, individual files are encrypted with separate keys. Although the file system encryption protects virtually all the files in a disk drive, other data outside the file system are omitted. Full disk encryption secures data on disk drives with a single symmetric key. Full disk encryption protects data in all areas of the disk drive, including areas outside the file system. Such data are the hidden files, swap files, file metadata, temporary files and caches, registry files, and boot sector data [22][23][24][25].

The preservation and acquisition of an encrypted disk drive can be tricky, depending on the power state, level, and type of encryption used – hardware-based or software-based [5]. It may be easier to preserve and acquire a file system-encrypted disk drive than a fully-encrypted disk drive in the powered off state. Likewise, the acquisition of a software-implemented and fully-encrypted disk drive may be easier than a hardware-implemented and file system-encrypted disk drive. Acquisition may be totally impossible in cases where disk is not accessible [8].

The examination and analysis phases of digital forensics investigation suffer the most from encryption technology. Reference [7] explained that there may be a possibility to recover an encrypted data, but it is often impossible to process the data. An examiner's tool needs access to read the contents of the encrypted data to be processed. However, they downplayed the threat posed to digital forensics by stating the possibility to circumvent encryption technology, even though it may be time-consuming and luck-dependent.

## IV. DATA ENCRYPTION TOOLS AND KNOWN FORENSICS MANOEUVRING METHODS

There are numerous data encryption solutions for disk drives. Each solution addresses data protection and privacy requirements using different methods. Some encryption solutions are compatible with particular operating systems, unlike others who are portable. They protect data at different levels and employ different key management and authentication methods. Most encryption solutions are implemented as software, but hardware-based solutions are preferred by some organizations [25]. Here is a list of popular disk drive encryption solutions: Microsoft's BitLocker, Symantec's PGP, Apple's FileVault, WinMagic's SecureDoc, IronKey's D200, RSA Data Security's RSA SecurPC, and McAfee's Endpoint [25][26][27]. However, the TrueCrypt solution has been briefly examined in this paper because of its immensely controversial reputation and gross utilization [8][9][22].

Although investigators are not entirely incapacitated by data encryption, sometimes it is down to the legal system to help get access to evidential data [7]. They stated that perpetrators used to employ the file-system encryption because they are concerned only about the data that are incriminating in their own opinions. The other areas left unencrypted usually contain sufficient evidence to prosecute them.

However, the threat posed by a full disk encryption solution is more detrimental to digital forensics processes. But as the saying goes that "no machine is 100% efficient", these encryption solutions have some exploitable vulnerabilities.

The encryption status seizes to hold for all data on the entire disk from the point the symmetric key has been accepted by the system until it is shut down. Deductively, data becomes accessible and inaccessible when the system is powered on and off respectively [7][25]. Digital forensics investigators need to execute an authorized and well-planned "search and seizure", with the aim of catching the perpetrator unawares while his system is running.

Another way to circumvent the threat is the traditional search for the encryption key [25]. There is a possibility, no matter how unlikely, that the key to decrypt the disk drive is written on a notepad or stored in USB drive somewhere at the scene.

Advanced memory-based procedures are also used to overcome the encryption threat. The concept of the full disk encryption that decrypts and makes data available for use is a memory function. The encryption key is stored in the memory the first time it was supplied. It remains in the memory and is used to automatically decrypt required data until the system is powered off. Various techniques can be used to retrieve the encryption key from the memory. RAMs hold data for few seconds to minutes – extendable by keeping the RAM cooled – without power. They can then be accessed through dedicated tools, such as MoonSol's Windows Memory, GMG Systems'

KnTList, Passware or F-Response. However, tools such as Wiebetech's HotPlug can be used to transfer the running system to a back-up power supply in case required expertise is not available and seized system needs to be moved to a laboratory [5].

### A. The TrueCrypt Scenario

TrueCrypt was first released in 2004, as a privacy-enforcing solution by the TrueCrypt Foundation. It is a very sturdy, free and open-source software solution, which allows intentional or accidental (sudden power outage), partial or full encryption and decryption processes, without compromising its data security ability. Several encryption algorithms used by TrueCrypt to encrypt data include Advanced Encryption Standard (AES), TwoFish, Serpent, AES-Twofish, Serpent-AES, AES-Twofish-Serpent, Twofish-Serpent, and Serpent-Twofish-AES. The PIPEMD-160, SHA-512 or Whirlpool hash algorithms, with a Random Number Generator, are used to create key-files for stronger security. TrueCrypt is capable of running in a portable mode, without being installed on the target disk drive. It supports smart cards and tokens for added security level and provides hot-keys to perform encryption tasks swiftly. It encrypts data in a partition/drive as a whole, or in a file-hosted container [5][9][22][24][27][28][29].

#### 1) Challenges Posed by TrueCrypt

TrueCrypt employs both the file-system and full-disk encryption methods. This ensures that all data, including registry entries, swap space, file metadata, temporary files, and hidden data are protected from unauthorized access. This leaves investigators with the options to plan a surprise access to a running system, search for possibly exposed encryption key, or use advanced memory-fetching techniques for keys or data [9][24][29].

TrueCrypt is an on-the-fly-encryption (OTFE)-based software. This concept ensures that only the data required by the user is available for access. All data are encrypted from the onset. The required data is copied into the memory and decrypted for use there. As such, the data on the disk drive remain encrypted. Once the data being processed by the user has changed and need to be saved to disk, TrueCrypt encrypts and saves it to the disk. Further, restricting investigators' access methods to those highlighted earlier [24].

Plausible deniability is a feature confidently boasted by TrueCrypt. The encrypted volume/drive appears as a drive with random data. A suspicious reaction to the randomness can be quashed with an explanation of a normal wipe process that fills drive with random data. This plausible deniability is possible because TrueCrypt does not have any signature that would be otherwise found in the drive's partition table [5][24][28][29].

It also allows volumes and operating systems to be hidden inside a visible TrueCrypt volume. The hidden volume is encrypted by a different key than the volume which contains it. It resides within the illusional random data created by the encryption of the visible volume. The user choses the volume to mount for use by the encryption key he supplies. If the legal system forces him to disclose his encryption key, he can disclose the key for the encrypted volume alone. The hidden volume, which will usually contain evidential information, remains oblivious to investigators [24][30].

A case example was a suspected terrorist, whose laptop was turned on and had the TrueCrypt mount window displayed. After initial refusal to disclose the encryption key, a High Court order requested he did. The case was dismissed later on, as no evidence could be gathered by the investigation/prosecution team against him (5). It is suffice to say that the suspected terrorist might have supplied the outer encrypted volume's key rather than the hidden volume's key. The plausible deniability feature of TrueCrypt proves to be a higher defeat to the three get-around digital forensics methods currently being used. Since, the TrueCrypt encryption solution is a widely available free software, experienced cyber-criminals will continue to use the features to avoid being prosecuted.

The following section looks at possible counter-actions to the robustness offered by encryption solutions for disk drives.

## V. MUTED SOLUTIONS TO DATA ENCRYPTION THREATS

Reference [8] highlighted two possible solutions to the forensics threats posed by inaccessible information containers. The first was the collaboration of digital forensics experts and storage device manufacturers to develop and implement a standard back-door across all storage devices.

This would seem the perfect solution, but Forte thought it was an excessive ask and "very unlikely to succeed". Three years on, there has not been any such collaboration or development. Forte pointed that the other solution involves a cooperation of the parties involved in an incident towards the adjustment of evidence preservation and analysis methodologies. He however admitted that the latter is only possible in e-discovery processes where both parties may be willing to cooperate rather than cyber-crime incidents whose subjects are often reluctant. Undecided about what solution was more feasible, even though the former seemed more effective, [8] believed there was enough time to find solutions before encrypted storage containers become pervasive, and as sort lightened the push for the former solution.

Reference [24] itself, highlighted that physical or remote (by malware or rootkit) access to such truecrypt-protected computer can compromise the encryption. The access may be used to install keyloggers and memory capturing software or hardware devices. These devices can obtain the encryption keys and passwords, as well as the unencrypted data which could be decrypted using acquired keys. Kleissner's "stoned" bootkit is a particular example of rootkit that can give such access to a target computer. It infects and controls the master boot record of the truecrypt program, and consequently allows the user to bypass the full volume encryption feature of truecrypt to access data resident on the computer [31]. However, the law enforcement agencies would be breaching the privacy rights of such individuals if physical or remote access to their computer is gained without their consent. A situation where law enforcement agencies have legal right to physically and remotely access a target computer is the sex offender monitoring case, in which the offender has actually been convicted of the crime. Unfortunately, a yet-to-be-convicted suspect that has actually committed a cybercrime will ordinarily not give such approval. He may also protect himself with the self-incrimination legal clause. Yet again, privacy needs have rendered another potentially-viable solution illegal.

## VI. CONCLUSION

The effectiveness of data encryption as a mechanism for enforcing information privacy is massive. This is evident by the reported widespread use of various data encryption solutions at the organizational and individual levels. However, its huge success for data access restriction has been a threat for digital forensics processes over the years. Cyber-criminals have been exploiting the information confidentiality ability of data encryption solutions, to restrict digital forensics investigators' accesses to potential evidence. The ubiquitous availability, inexpensive cost and easy implementation of encryption solutions enhance the threats posed to digital forensics processes. Investigators sometimes get around the encryption challenge through careful and thoughtful planning of search and seizure, thorough search for exposed encryption keys, and advanced in-memory data retrieval techniques. Yet, a minimum of 60% of computer incidents involving data-encryption end up unprosecutable.

The TrueCrypt software went even further by providing users with plausible deniability and non-repudiation abilities. This makes digital forensics investigations of encrypted disk drives harder and less feasible. Consequently, this undesired situation constitutes an indirect reason for the rise in occurrence of computer incidents. As much as data encryption helps offenders get away from being caught, the necessity for data privacy and security cannot be sacrificed for digital forensics. Unfortunately, the only digital forensics solution to a threatening information security solution will have to be unanimously considered by disk drive manufacturers. There should be a technology that will provide a backdoor for digital forensics investigators to gain access to the most securely encrypted disk drives. However, there will have to be a restriction to the distribution of such technology when it comes to existence. This is to avoid its abuse by non-law enforcement practitioners (and potential computer criminals) to illegally access target data.

### ACKNOWLEDGMENT

The encouragement received from Dr. Shao Ying Zhu was quite appreciated by the author.